\documentclass{iau}
\usepackage{graphicx}

\newcommand{\msun}{M_\odot}
\newcommand{\uconv}{u_{\rm conv}}

\title[Do Magnetic Fields Actually Inflate Low-Mass Stars?] 
{Do Magnetic Fields Actually Inflate Low-Mass Stars?}

\author[Feiden \& Chaboyer]   
{Gregory A. Feiden$^{1, 2}$ \and Brian Chaboyer$^{2}$}

\affiliation{$^1$Dept. of Physics \& Astronomy, Uppsala 
                 University, Box 516, Uppsala 751 20, Sweden.
                 \\ [\affilskip]
             $^2$Dept. of Physics \& Astronomy, Dartmouth College, 6127 
                 Wilder Laboratory, Hanover, NH 03755, USA.
                 \\
                 email: {\tt gregory.a.feiden.gr@dartmouth.edu}
             }

\pubyear{2013}
\volume{302}  
\pagerange{1--4}
\jname{Magnetic Fields Throughout Stellar Evolution}
\editors{M. Jardine, P. Petit \& H. Spruit, eds.}

\begin{document}
\maketitle

\begin{abstract}
Magnetic fields have been hypothesized to inflate the radii of low-mass 
stars---defined as less than 0.8\,$\msun$---in detached eclipsing 
binaries (DEBs). We evaluate this hypothesis using the magnetic Dartmouth 
stellar evolution code. Results suggest that magnetic suppression of thermal 
convection can inflate low-mass stars that possess a radiative core and 
convective outer envelope. A scaling relation between X-ray luminosity and 
surface magnetic flux indicates that model surface magnetic field strength
predictions are consistent with
observations. This supports the notion that magnetic fields may be inflating 
these stars. However, magnetic models are unable to reproduce radii 
of fully convective stars in DEBs. Instead, we propose that model 
discrepancies below the fully convective boundary are related to metallicity.
\end{abstract}

\firstsection 
\section{Introduction}
It has been well-documented over the past decade that stellar evolution models
are unable to accurately predict radii and effective temperatures---so 
called ``fundamental properties''---of low-mass stars ($M < 0.8\,\msun$) in 
detached eclipsing binaries (DEBs; see, e.g., \cite{Ribas2006,FC12a}). 
Model radii have been shown to be too small and effective temperatures too 
hot, particularly in the most well-studied systems. Since stellar evolution 
models are heavily 
used to aid with the interpretation of observational data, it is crucial
that these modeling errors be addressed. Magnetic fields, maintained by
spin-orbit synchronization of DEB components, have been hypothesized
to be the culprit. Magnetic activity indicators appear to correlate with
radius discrepancies, supporting this hypothesis (e.g., \cite{Lopez2007}).
We aim to test this hypothesis using the recently developed magnetic Dartmouth 
stellar evolution code.

\section{Method}
We use models generated as a part of the Dartmouth Magnetic Evolutionary 
Stellar Tracks and Relations program (DMESTAR; \cite{FC12b}, 2013a) to test whether 
model radii may be inflated by interactions between a magnetic field and 
thermal convection. Two techniques are used to incorporate magneto-convection
in the models: (1) stabilization of convection, and (2) 
inhibition of convective efficiency (\cite{FC13a}). Method one alters 
the Schwarzschild convective stability criterion by assuming the magnetic 
field is in equilibrium 
with the surrounding gas. An upper limit to the magnetic field strength 
occurs when the magnetic field is in thermal equipartition with the gas, 
approximately when the magnetic pressure is equal to the gas pressure. 
Method two instead assumes that the energy
required to create the magnetic field is drawn from the kinetic energy
of convective flows. Therefore the magnetic field strength has an upper
limit determined by equipartition with convective flows,
$B_{\rm eq} = (4\pi\rho\uconv^2)^{1/2}$.

Standard stellar evolution models (i.e., non-magnetic models) were run for
both components of DEB systems to assess the level of radius inflation 
required of our magnetic models. A series of magnetic models were then
computed with varying magnetic field strengths, using the magneto-convection
techniques outlined above, until the model fundamental properties agreed
with observationally determined values. We performed this procedure for
several well-studied DEB systems to ascertain whether results were robust. 

\section{Results}
Effects of magneto-convection on fundamental stellar properties predicted
by stellar evolution models may be summarized for two different low-mass 
stellar populations: stars that have a radiative core with a convective 
outer envelope and fully convective stars. We will focus on
model radius predictions, as stellar radii are more reliably measured from
observations than effective temperatures.

\begin{figure}[t]
    \centering
    \includegraphics[scale=0.34]{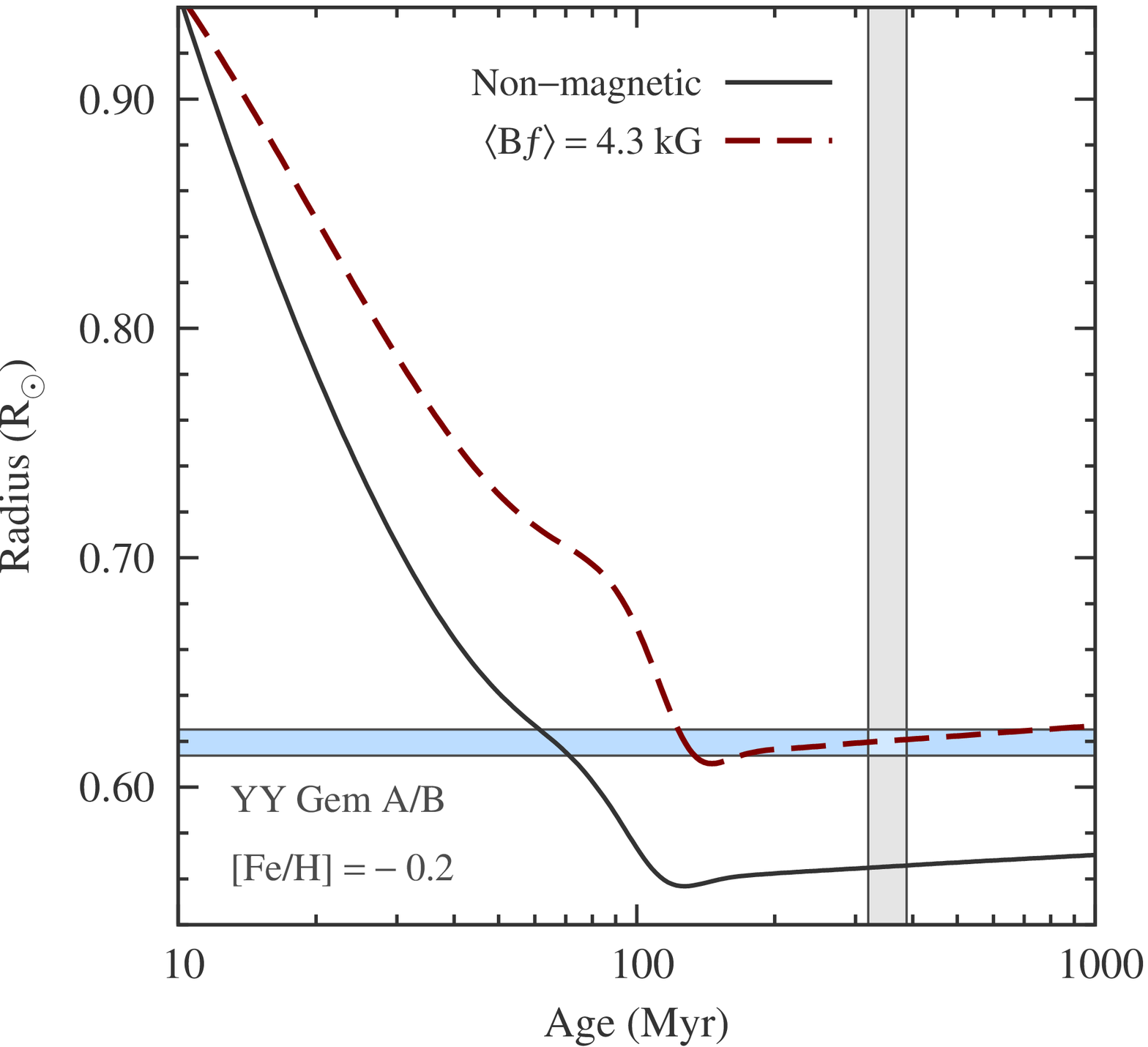}
    \qquad
    \includegraphics[scale=0.34]{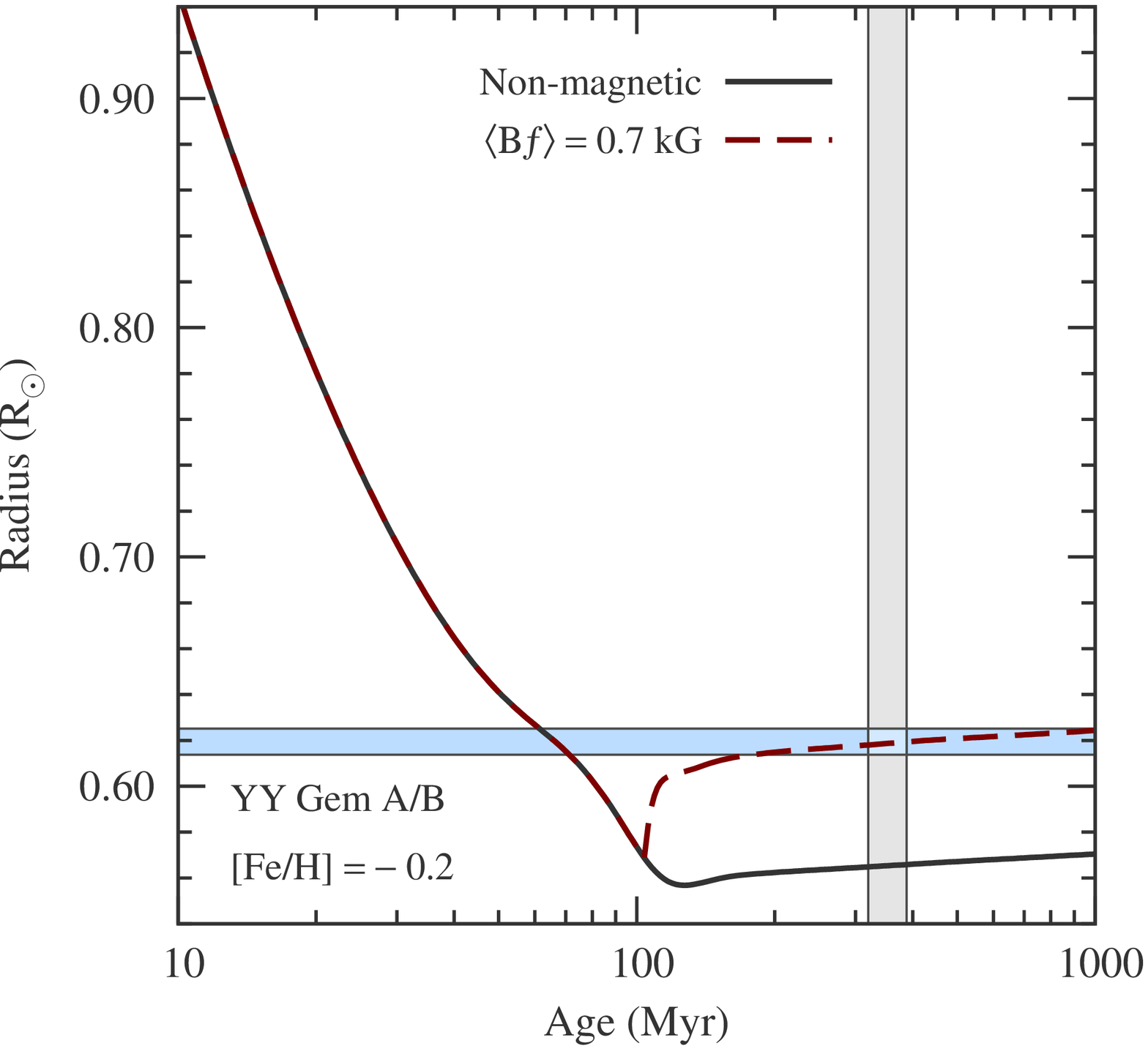}
    \caption{Radius evolution of standard (black, solid line) and magnetic 
             (maroon, dashed line) models of the equal-mass DEB YY Gem. 
             ({\it Left}) Magnetic
             model using stabilization of convection introduced at 10\,Myr. 
             ({\it Right})
             Magnetic model invoking inhibition of convective efficiency
             initialized at 100\,Myr. The perturbation age has 
             no effect on results along the main sequence.
             The blue horizontal shaded region indicates the observed 
             radius with associated $1\sigma$ uncertainties and the 
             vertical grey region highlights the estimated age of the 
             system.}
    \label{fig:yy_gem}
\end{figure}

The influence of magneto-convection on partially convective stars is demonstrated
in Figure\,\ref{fig:yy_gem}. The only significant difference between the 
two panels is the adopted magneto-convection technique with stabilization 
of convection used in the left panel and inhibition of convective efficiency 
in the right panel. It is clear that accounting for magneto-convection can 
inflate model radius predictions at a level required to reconcile models 
with observations. In general, we observe that both magneto-convection 
techniques provide a qualitatively correct solution for partially
convective stars. However, the two magneto-convection techniques predict 
significantly different surface magnetic field strengths. Inhibition of 
convective efficiency typically requires weaker surface magnetic field 
strengths---by roughly a factor of 5---than stabilization of convection.
For example, Figure\,\ref{fig:yy_gem} shows that stabilization of convection 
(left panel) requires a 4.3\,kG surface magnetic field while inhibition 
of convective efficiency (right panel) requires a 0.7\,kG surface magnetic 
field to correct models of YY Gem to observations. In both cases the peak 
interior magnetic field strengths are roughly on the order of 1 -- 10\,kG.

\begin{figure}[t]
    \centering
    \includegraphics[scale=0.35]{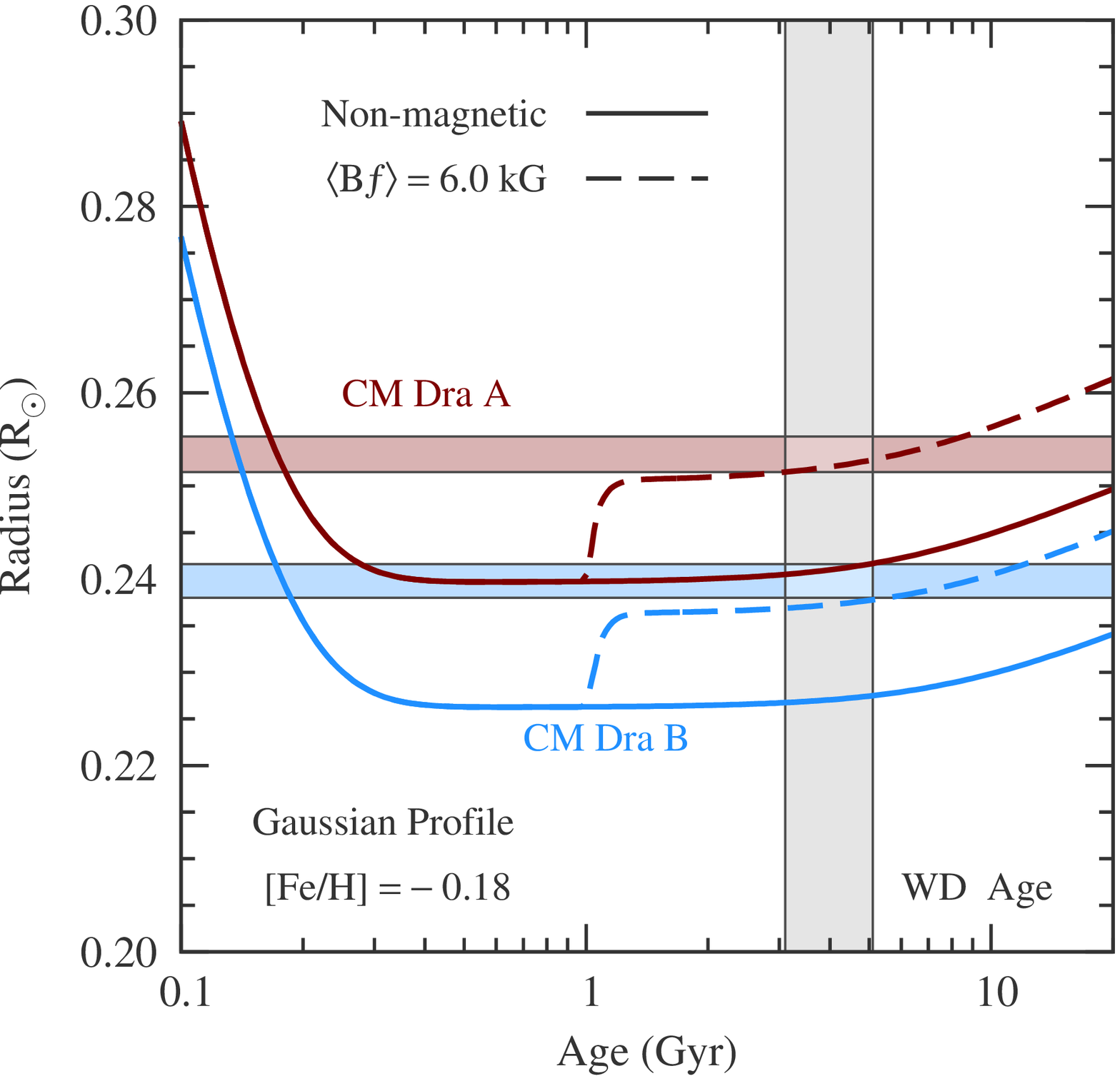}
    \qquad
    \includegraphics[scale=0.35]{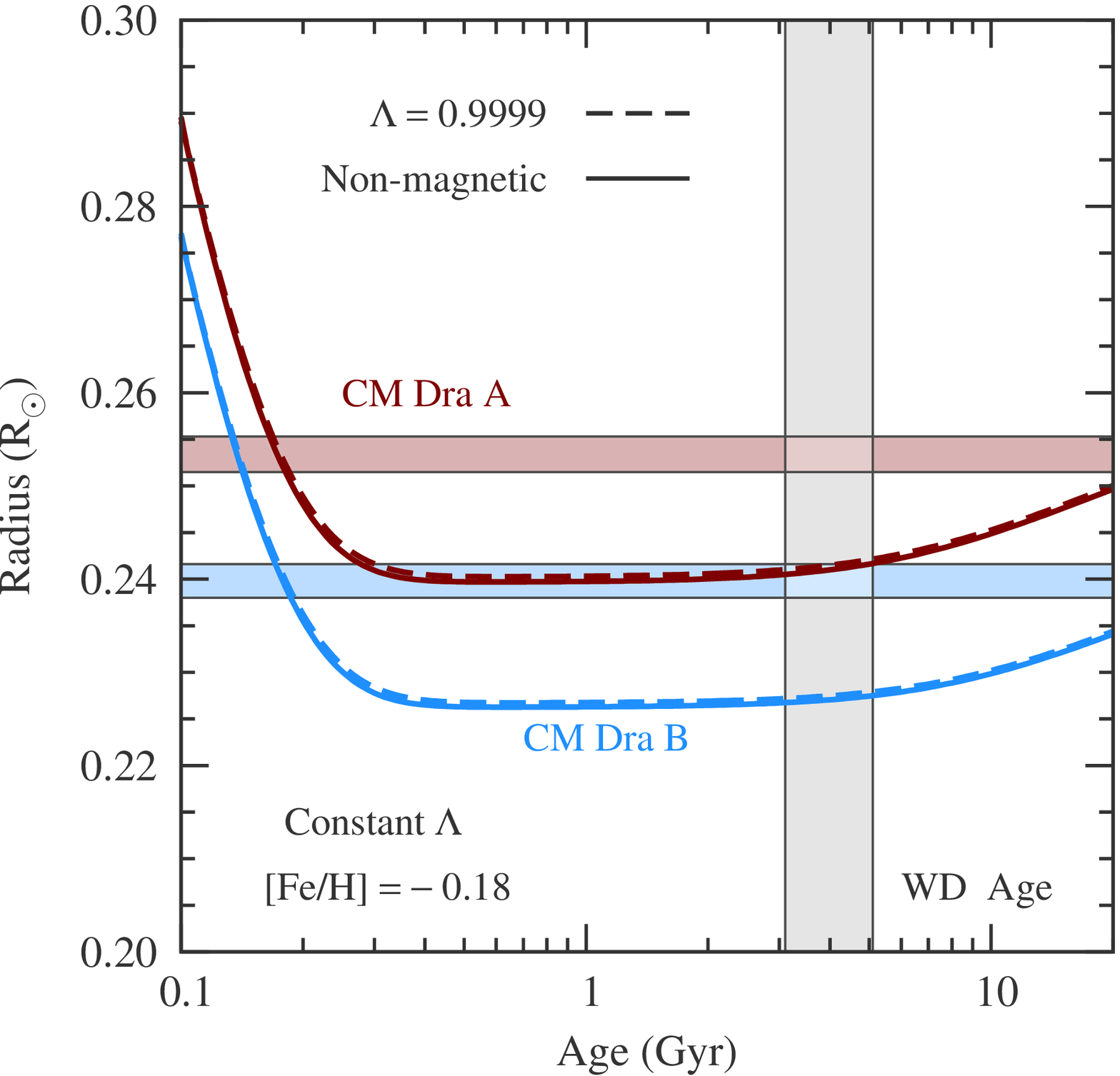}
    \caption{Same as Figure \ref{fig:yy_gem}, but for the stars of
             CM Dra. (\textit{Left}) Models invoking the stabilization 
             of convection initialized at 1\,Gyr. (\textit{Right}) Models 
             that have inhibited
             convective efficiency initialized at 100\,Myr.}
     \label{fig:cm_dra}
\end{figure}

Results are quite different for fully convective stars, as illustrated with
a representative system (CM Draconis) in Figure\,\ref{fig:cm_dra}. The 
left panel shows that stabilization of convection is sufficient to reconcile 
model radii with observations. Models presented in the left panel of 
Figure\,\ref{fig:cm_dra} have a surface magnetic field strength of 6.0\,kG
with a peak interior magnetic field strength of roughly 50\,MG. Inhibiting 
convective efficiency, however, does not radically alter model predictions.
Models in the right panel of Figure\,\ref{fig:cm_dra} have magnetic field
strengths equal to 99.99\% of the equipartition value---roughly 3\,kG at
the surface and 50\,kG deep within the star. What is not apparent from
Figure\,\ref{fig:cm_dra} is that we adjusted the interior magnetic field 
strength within the models invoking stabilization of convection to provide
greater radius inflation. Fully convective models are sensitive to
the deep interior magnetic field strength, which can be set arbitrarily
to achieve the desired inflation.

\section{Discussion}
Introducing magneto-convection into stellar models appears to provide at
least a qualitative solution to the problem of inflated low-mass stars in
DEBs. What must be addressed is whether the magnetic field properties (surface
and interior field strengths) are physically realistic. Using a scaling 
relation between stellar coronal X-ray luminosity and surface magnetic 
flux (\cite{FC13a}), we find that models invoking the stabilization of 
convection require surface magnetic field strengths that are likely too
strong. This is particularly evident for partially convective stars where
estimated surface magnetic field strengths are too strong by about a factor
of 5. However, models using inhibition of convective efficiency predict
surface magnetic field strengths consistent with X-ray luminosity estimates.
Assuming a ``turbulent dynamo'' is primarily responsible for the generation
of magnetic fields in low-mass partially convective stars, it appears 
plausible that magnetic fields are inflating stellar radii.

Surface magnetic field strengths for fully convective stars invoking stabilization
of convection appear plausible based on X-ray luminosity estimates. Models
using inhibition of convective efficiency, while providing the best agreement
with observed surface magnetic field strengths (\cite{Reiners2012}),
fail to reproduce observed stellar properties. Recall, we mentioned that
the deep interior field strength in fully convective stars was important. 
Magnetic field
strengths must be nearly 50\,MG to influence the structure of low-mass
fully convective stars. This appears too strong to be physically
plausible. A turbulent dynamo mechanism, such as that operating in fully
convective stars, cannot produce a magnetic field strength of such magnitude.
Instead, a field of that strength must be of primordial origin, which 
appears inconsistent with estimated diffusion timescales and additional 
observational evidence (\cite{FC13b}). Therefore, it seems unlikely, given
current evidence, that magnetic fields are responsible for inflating fully
convective stars.

\begin{figure}
    \centering
    \includegraphics[scale=0.35]{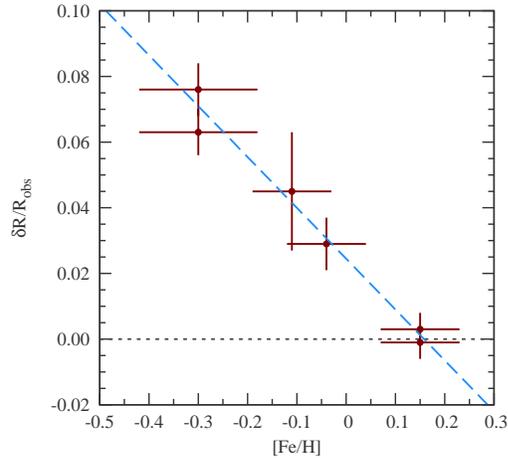}
    \caption{Relative radius error between stellar evolution models
             and observations of fully convective stars plotted against
             observationally determined metallicities.}
    \label{fig:feh_resid}
\end{figure}

Ultimately, we must then view the result for partially convective stars
with some skepticism. Further theoretical modeling of magneto-convection
and additional observational constraints on properties like metallicity,
surface magnetic flux, and star spot coverage are required. Yet, it is 
not too early to begin searching for other explanations. Among fully convective
stars in DEBs, for instance, we find a strong anti-correlation between the 
level of radius inflation of real stars and estimated stellar metallicities 
(see Figure\,\ref{fig:feh_resid}). At the moment, we have no convincing 
explanation for this anti-correlation, just that we see it in the DEB data
and hints of it in interferometric data (\cite{Boyajian2012}). \\

{\it This work was supported by NSF grant AST-0908345 and the William
H.\,Neukom 1964 Institute for Computational Science at Dartmouth College.}

%
%

\end{document}